\documentclass[12pt]{article}
\usepackage{graphicx}

\usepackage{booktabs}
\usepackage{verbatim}

\usepackage{hyperref}
\usepackage{multirow, makecell}
\usepackage{siunitx}

\newcommand{\kL}{$\rm k_1$}
\newcommand{\dataA}{$\mathcal{D}^{\,500}_{\,1.0}$ }
\newcommand{\dataB}{$\mathcal{D}^{\,500}_{\,0.2}$ }
\newcommand{\dataC}{$\mathcal{D}^{\,250}_{\,1.0}$ }

\date{\today}
\title{\bf Analysis of GSI HADES beamline configuration space}
\author{Mariusz Sapinski\thanks{m.sapinski@gsi.de},\\ 
GSI Helmholtzzentrum für Schwerionenforschung, Darmstadt, Germany \\
Dominik Vilsmeier, \\
Goethe Universit\"at Frankfurt, Frankfurt, Germany \\
} 

\begin{document}
\maketitle
\thispagestyle{empty}

\begin{abstract}
The High-Energy Beam Transport system of GSI Campus is a complex system of interconnected beamlines between SIS18 synchrotron and multiple experimental stations and storage rings. 
This system is designed to be flexible and each experimental station can be reached by several beam paths.
The quadrupole settings must fulfill constraints due to the beamline acceptance, limitations of the installed magnets 
and experimental requirements. The number of degrees of freedom is large 
and leaves operators with a vast number of possible optics configurations.
The choice of a particular configuration is often a matter of compromises, very specific requirements, operational experience and personal style.
Once a satisfying configuration is found, the space of possible solutions is usually not explored any further.
%All those requirements limit the configuration space of possible
%beamline settings in a complex way, which is not easy to visualize.
Here the Monte Carlo method is used to probe the space of all possible beamline optics configurations. 
The focus is on a particular section, the HADES beamline, because this is one of the most demanding lines due to high beam intensities required by the experiment. The presented analysis gives an overview of the possible settings, insight into beamline potential and flexibility and supports 
the choice of an optimal operational setting.

%he strength of magnets (mainly quadrupoles) %istuned according to the requirements
%of the experiment or accelerator accepting the beam, but also following the acceptance of the beamline.
%The optics should allow for an efficient and precise beam transport to the target or to another machine. 
%For a typical, relatively long beamline, there exist many good solutions
%to the problem of efficient transporting of the beam to the target. 
\end{abstract}

\thispagestyle{empty}

%\tableofcontents

\newpage

%%%%%%%%%%%%%%%%%%%%%%%%%%%%%%%%%%%%%%%%%%%%%%%%%%%%%%%%%%%%%%%%%%%%%%%%%%%%
\section{Introduction}

A transfer line is a part of an accelerator facility without accelerating cavities, used
to transport the beam between accelerators or from an accelerator to experiments.
The main constituents of a beamline are dipoles, quadrupoles and steerer magnets. Other types of magnets are rarely used.
Dipole settings are defined by the geometry of the beamline.
Quadrupoles, together with beam properties at the entrance of the beamline, define what is called \textit{beam optics}.

\begin{figure}[htb]
   \centering
   \includegraphics[width=.55\textwidth]{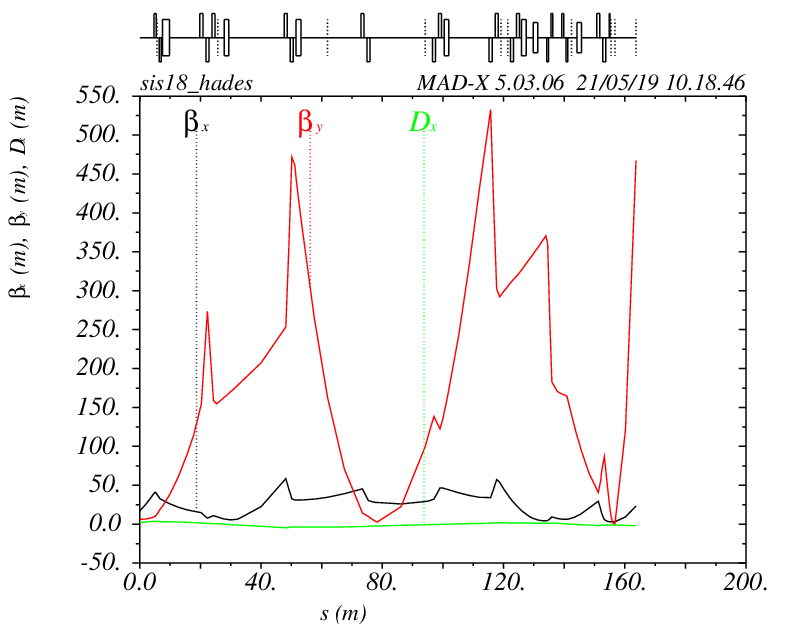} 
   \caption{Example of ion optics for HADES beamline used during the beam time in 2019. The experimental target is located at \SI{158}{\meter}.}
   \label{fig:HADES_optics2019}
\end{figure}

In GSI, the network of beamlines starting at SIS18 heavy ion synchrotron is called HEST (from German "Hochenergie-Strahlführung") \cite{HESTweb}.
The beamlines bifurcate, cross and reunite in a particular, complex pattern adapted to the experiments.
HADES \cite{HADES} is one of the largest experiments and 
it is placed at the end of an about 160 meter long beamline starting at the magnetic septum of SIS18. 
The beamline contains 21 individually powered quadrupoles and two active dipoles tilted by \SI{21.7}{\degree} to bring the beam to the elevated position of the experimental area \cite{Sapinski:2019mvl, Sapinski:2017}. HADES is designed to work in two main modes, using either primary or secondary particles. 
Here the first case is studied, in which the beam is focused on an internal target inside the experiment. In both cases HADES accepts slowly extracted beams which are realized by incrementally changing the tune towards third order resonance.
The required beam spot size is about 
$\sigma=\SI{0.4}{\milli\meter}$ \cite{Rost2019}.  
%\SI{0.2-0.5}{\milli\meter}. 
Only about \SI{1}{\percent} of ions interact with the target and the rest of the beam is dumped
downstream of the experiment.
A typical ion optics used in operation, called \textit{BEAMTIME2019}, is visualized in \hbox{Figure \ref{fig:HADES_optics2019}}.

Setting up a beamline optics involves fulfillment of various constraints. Typically the beam should be transported with maximum transmission
and focused on the experimental target.
%The quadrupole magnets focus the beam in order to fit the beam envelope into the beamline aperture to minimize beam losses.
%Because the vertical aperture is usually small inside the dipole magnets, they are important points to consider when preparing the optics.
%In the phase of the beamline design, the quadrupole magnets should be foreseen upstream of dipoles to provide enough focusing strength.
%Over-focusing, which is the situation when the focus point after a quadrupole is before the next quadrupole, should be avoided.
%In order to preserve the beam emittance the maximum focus should be foreseen at the intermediate targets and stripper foils.
These constraints can be fulfilled by various quadrupole configurations.
Therefore, an analysis of the set of all possible configurations can provide interesting insights about the beamline capabilities.

The relevant configuration space has 21 dimensions corresponding to the 21 quadrupoles. 
Because of its high dimensionality the volume of this space is extremely large and thus it is not possible to generate and compute all possible configurations. A randomly sampled subset of configurations can however be chosen in order to represent the full population.

This paper is structured as follows: in Section \ref{sec:Sample} the probing of configuration space and generation of data sets is described, in Section \ref{sec:Optics} the general properties of optics functions are discussed, in Section \ref{sec:ConfigurationSpace} the configuration space is analyzed and in Section \ref{sec:DimensionalityReduction} the potential for dimensionality reduction is investigated. In Section \ref{sec:Microstructure} the microstructure of the configuration space is investigated.
Section \ref{sec:PCA} presents Principal Component Analysis of the \kL-values. In Section \ref{sec:Clustering} a grouping of configurations is discussed. Section \ref{sec:Robustness} discusses stability of the optics, considering quadrupoles gradient errors as well as a shifting of Twiss parameters at the entrance of the beamline. Finally, the last Section contains a discussion on the choices of the operational ion optics in view of previous results.

%%%%%%%%%%%%%%%%%%%%%%%%%%%%%%%%%%%%%%%%%%%%%%%%%%%%%%%%%%%%%%%%%%%%%%%%%%%%
\section{Samples}
\label{sec:Sample}

Three data sets with \num{10000} configurations each have been generated. They use randomly sampled \kL-values for the 21 quadrupoles which are used as starting points for a subsequent MADX \cite{MADX} matching procedure.
% the Levenberg-Marquardt algorithm (LMDIF) was used. <details on LMDIF come later>

%The beamline has 21 quadrupoles so this is the number of parameters.
The three data sets differ in the matching constraints on the beta function which are described in Table \ref{tab:samples}. In addition to these constraints the $\beta_{h,v}$ on the beam dump, downstream of the 
experimental target, is constrained to less than \SI{3000}{\meter}. 
The Levenberg-Marquardt algorithm \cite{lmdiff}, which is implemented as \textit{LMDIF} in MADX, is used as the matching method with the maximum number of function evaluations set to \num{1d7} and a tolerance of \num{1d-16}.
Only \SI{0.03}{\percent} of the randomly chosen starting points converged during the matching procedure. Table \ref{tab:finish-reasons} contains an overview of the various termination reasons of the LMDIF algorithm. The data sets contain only successfully converged configurations for the further analysis.
The further analysis is concentrated around the \dataA data set.

\begin{table}[!hbt]
   \centering
   \caption{Summary of data sets and associated matching constraints.}
   \begin{tabular}{l|cc|}
       \toprule
       & \multicolumn{2}{|c|}{\textbf{Constraints}} \\
       \textbf{data set} & \textbf{beamline} & \textbf{target}  \\
       \midrule
           \dataA   &   $\beta_{h,v} < \SI{500}{\meter}$   &   $\beta_{h,v} < \SI{1.0}{\meter}$   \\
           \dataB   &   $\beta_{h,v} < \SI{500}{\meter}$   &   $\beta_{h,v} < \SI{0.2}{\meter}$   \\
           \dataC   &   $\beta_{h,v} < \SI{250}{\meter}$   &   $\beta_{h,v} < \SI{1.0}{\meter}$   \\
       \bottomrule
   \end{tabular}
   \label{tab:samples}
\end{table}

\begin{table}[!hbt]
   \centering
   \caption{Overview of the various LMDIF termination reasons that occurred during probing of the configuration space. Here "unstable" means that the six-dimensional orbit vector grew too large along the beamline during the matching process.}
   \begin{tabular}{lr}
        \toprule
        \textbf{Reason} & \textbf{Fraction} \\
        \midrule
        unstable                     &  \SI{93.8370}{\percent} \\
        converged without success    &  \SI{ 3.2336}{\percent} \\
        variables too close to limit &  \SI{ 2.8957}{\percent} \\
        converged successfully       &  \SI{ 0.0336}{\percent} \\
        call limit                   &  \SI{ 0.0002}{\percent} \\
        \bottomrule
        \end{tabular}
   \label{tab:finish-reasons}
\end{table}

The matching constraints lead to strong limitations and interrelations with respect to the \kL-values. 
The results of the matching procedure depend not only on the matching conditions but also on the used algorithm.
The Levenberg-Marquardt minimization uses gradient descent and it stops as soon as all constraints are fulfilled. Thus it is possible that a given result could be further optimized towards even smaller values of the beta function but this possibility is not explored by the optimizer.

% The massive MADX simulations needed for this work were perfomed on GSI HPC cluster.
% The subsequent analysis of the properties of the set of good configurations is done using scikit-learn package \cite{sklearn}.

%%%%%%%%%%%%%%%%%%%%%%%%%%%%%%%%%%%%%%%%%%%%%%%%%%%%%%%%%%%%%%%%%%%%%%%%%%%%
\section{Optics properties}
\label{sec:Optics}

\begin{figure}[htb]
   \centering
   \includegraphics[width=.45\textwidth]{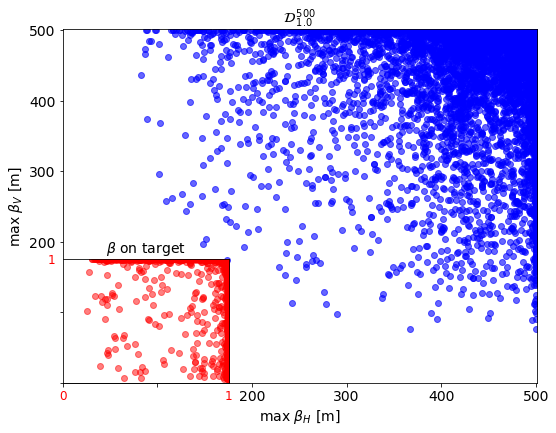} 
   \includegraphics[width=.45\textwidth]{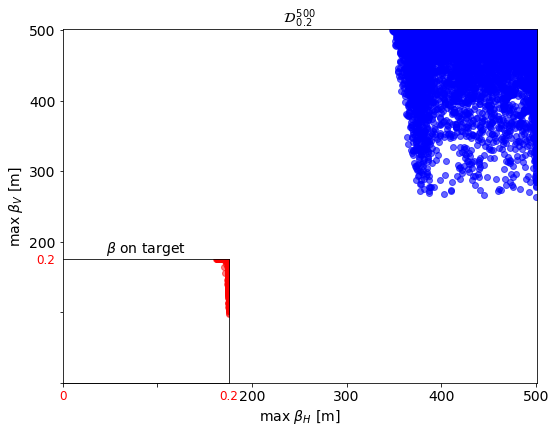} 
   \caption{Values of the final beta function along the beamline (maximum value, blue) and at the target location (red) for \dataA (left) and \dataB (right).}
   \label{fig:optics_properties}
\end{figure}

The minimization procedure stops as soon as all constraints are fulfilled, including cases where some of the constrained quantities end up well below their associated thresholds. This becomes apparent on the left plot of Figure \ref{fig:optics_properties}, where a large number of configurations
clearly surpass their constraints. Imposing stronger constraints results in a smaller possible margin and the constrained quantities remain closer to their threshold values, as can be seen from the right plot of Figure \ref{fig:optics_properties}. This suggests that the constraints of the \dataB case are close to the limits of the beamline.

The dispersion at the target is a relevant property of the optics as well which however was not constrained during the matching process. Vertical dispersion is small but non-zero due to the presence of tilted dipoles. Small horizontal dispersion is desirable because the beam momentum changes during the spill resulting in potential movement of the beam spot on target. In order to counteract this movement, the first steering dipoles after the extraction septum can be ramped during the spill and therefore zero dispersion at the target is not absolutely necessary but desirable nonetheless.

\begin{figure}[htb]
   \centering
   \includegraphics[width=.45\textwidth]{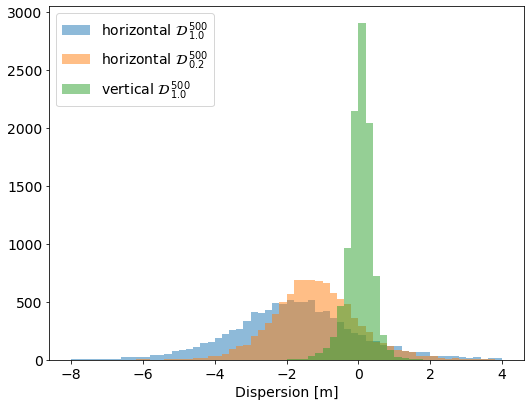} 
   \includegraphics[width=.45\textwidth]{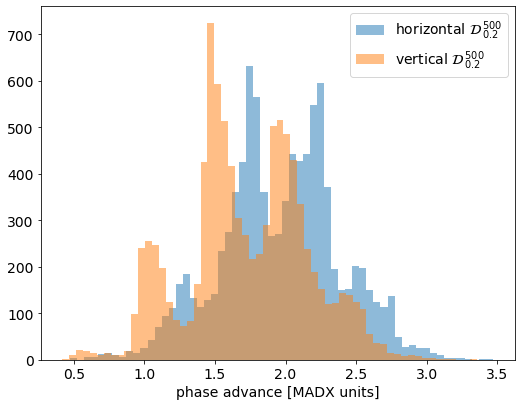} 
   \caption{Left: Distribution of horizontal and vertical dispersion on the target. Right: Distributions of the phase advance on the target plotted for \dataB sample; for \dataA a similar structure occurs but the peaks are less pronounced.}
   \label{fig:dispersion}
\end{figure}

The distribution of dispersion at the target is shown on the left plot of Figure 
\ref{fig:dispersion}. The right plot of this figure presents another interesting
aspect of the minimization procedure. The phase advance on the target tends to 
prefer values corresponding to certain angles (or 180-degree rotations). For instance the peaks for vertical phase advance are located around $\pi$, $1.5\pi$ and $2\pi$ for \dataB. For other data sets, the peaks are less pronounced. This shows that the whole beamline must be more precisely tuned when a stronger constraint on the target focusing is imposed.

%%%%%%%%%%%%%%%%%%%%%%%%%%%%%%%%%%%%%%%%%%%%%%%%%%%%%%%%%%%%%%%%%%%%%%%%%%%%%%%%
% jupyter 12
%\include{StatOptics_PhSpaceVol}
% new jupyter 3
\section{Configuration space}
\label{sec:ConfigurationSpace}

The configuration space spans across 21 dimensions defined by \kL-values
of the quadrupole magnets. The beamline contains two types of quadrupole magnets and their main properties are shown in Table \ref{tab:quads}.

\begin{table}[!hbt]
   \centering
   \caption{Quadrupole types installed on the beamline.}
   \begin{tabular}{lccc}
       \toprule
       \textbf{Type} & \textbf{Length} & $\mathbf{max\,|k_1|}$ & \textbf{Count} \\
       \midrule
           QPK & \SI{0.6}{\meter} & $\SI{0.37}{\per\square\meter}$ & 6 \\ 
           QPL & \SI{1.0}{\meter} & $\SI{0.60}{\per\square\meter}$ & 15 \\ 
      \bottomrule
   \end{tabular}
   \label{tab:quads}
\end{table}

The configuration space is a 21-orthotope with \num{2097152} vertices. The corresponding volume is
\begin{equation}
 V_{21} = \prod_{i=1}^{21} \max |k_{1,i}| = \SI{1.21d-6}{\meter^{-42}}
 \label{eq:hyperrvolume}
\end{equation}
The notion of volume in 21 dimensions is far from being intuitive and probably a better 
grasp is provided by the largest one-dimensional span between the edges of the configuration space.
This maximum Euclidean distance is \SI{2.49}{\per\square\meter}.

\begin{figure}[htb]
   \centering
   \includegraphics[width=.45\textwidth]{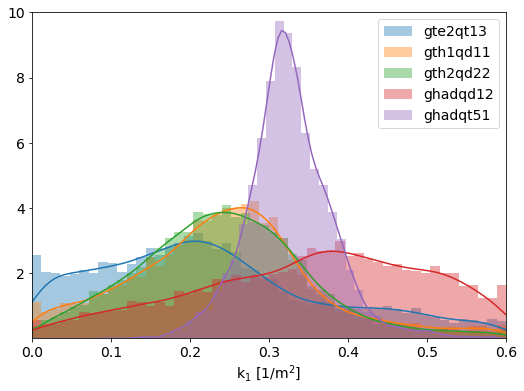} %
   \includegraphics[width=.45\textwidth]{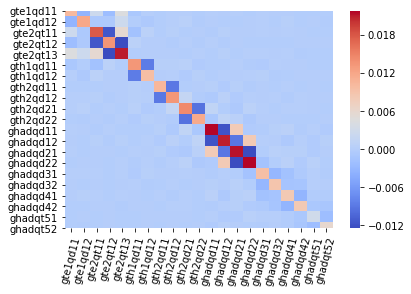} %
   \caption{Left: distribution of selected \kL-values selected as examples. 
   Right: covariance matrix for the \kL-values. The correlations between the quadrupole doublets are observed.}
   \label{fig:k1_dist}
\end{figure}

The available configuration space is strongly reduced by the constraints listed in Table \ref{tab:samples}. 
The distribution of resulting \kL-values is approximately continuous and it does not peak at the extremes of \kL-values,
%(however microstructure of the configuration space is not uniform, as it will be discussed later), 
as shown, for a selection of magnets, on the left plot of Figure \ref{fig:k1_dist}.
Therefore, the volume occupied by the 
converged configurations can be estimated from eigenvalues ($\rm\lambda_i$) of the covariance matrix using equation \ref{eq:phasesvol}. The eigenvalues of the covariance matrix represent the variances of the projection of the 21-dimensional distribution onto the eigenvectors. If these distributions are Gaussian, then width of the distribution $\sigma_{gauss} = \sqrt{\lambda}$. The distributions are not always Gaussian, especially for the eigenvectors with largest variance (see Section \ref{sec:PCA}), so the proposed estimation is an approximation. 
For non-Gaussian distributions Chebyshev's inequality can be used to estimate which percentage of configurations lie within a given range of variances, i.e. in the range $\pm 2\sqrt{\lambda}$ there should be between \SI{75}{\percent} (Chebyshev's inequality) and \SI{95}{\percent} (normal distribution) of configurations.
In addition, due to presence of substructures in the configuration space, this approach overestimates the actual space of valid parameters. This will be discussed further in Section \ref{sec:Microstructure}.

\begin{equation}
    V_{21} = 4^{21} \prod_{i=1}^{21} \sqrt{ \lambda_i}
    \label{eq:phasesvol}
\end{equation}

The covariance matrix is visualized on the right plot of Figure \ref{fig:k1_dist}. The quadrupole triplet and two doublets with largest values of matrix elements constitute the first two principal components (see Section \ref{sec:PCA}).

The volume of the configuration space which contains matched configurations is about $\rm 10^{-4}$ of the total volume.
As one can see from Table \ref{tab:phaspac}, constraining the beamline beta function by factor 2 (from 500 m to 250 m) decreases the configuration space volume by factor 10 and constraining the beta function on the target by factor 5 decreases the configuration space volume by factor 50.

\begin{table}[!hbt]
   \centering
   \caption{Configuration space volume calculated with Equation \ref{eq:phasesvol}.}
   \begin{tabular}{lc}
       \toprule
       \textbf{sample} & \textbf{Volume [$\rm m^{-42}$]}  \\
       \midrule
           \dataA   & $\rm 7.93\cdot 10^{-10}$   \\ 
           \dataC   & $\rm 6.29\cdot 10^{-11}$   \\ 
           \dataB   & $\rm 1.29\cdot 10^{-11}$  \\
       \bottomrule
   \end{tabular}
   \label{tab:phaspac}
\end{table}

\begin{figure}[htb]
   \centering
   \includegraphics[width=.45\textwidth]{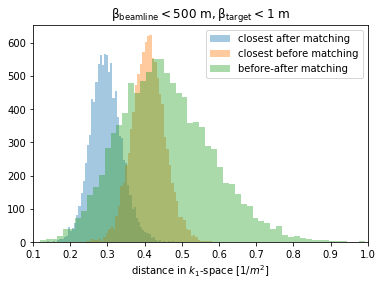} %
   \includegraphics[width=.45\textwidth]{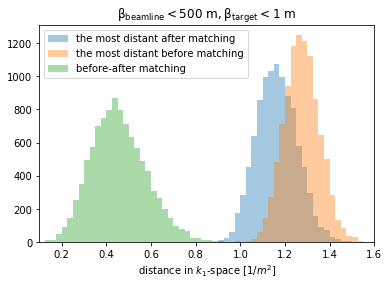} %
   \caption{Left: distribution of minimum distances between configurations. 
   Right: distribution of maximum distances between configurations.}
   \label{fig:k1_distances}
\end{figure}

The distribution of distances between initial and final configurations and distances to the closest and the most distant configuration are shown in Figure \ref{fig:k1_distances}. The average distance to the closest configuration before the matching procedure is \SI{0.41}{\per\square\meter} and drops to \SI{0.29}{\per\square\meter} for the converged configurations. This distance is never zero, the models do not overlap.
The average distance between corresponding initial and final configurations is
\SI{0.46}{\per\square\meter} which is significantly larger than the distance to the final nearest neighbour.
The maximum stretch of the reduced configuration space is only \SI{1.51}{\per\square\meter}.

The last characteristic length is related to the local size of a configuration, which will be discussed in more detail in Section \ref{sec:Microstructure}. It describes the local region which fulfills the matching conditions. 
Due to the high dimensionality of the problem, the precise shape of this region is difficult to estimate. 
An increase of sampling requires enormous computing power, not available for this study.

Two approaches are used in this study. In the first approach small, selected areas of the configuration space are sampled with very large granularity. In the second approach, for each configuration a set of 42 variations is created.
For each variation one of the \kL-values is changed by $\pm r_{21}$.
%Both approaches have limitations, for instance the last approach ignores the correlations between quadrupole doublets.

\begin{figure}[htb]
   \centering
   \includegraphics[width=.45\textwidth]{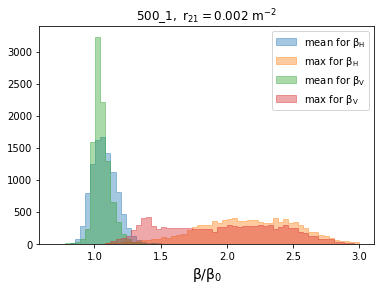} %
   \includegraphics[width=.45\textwidth]{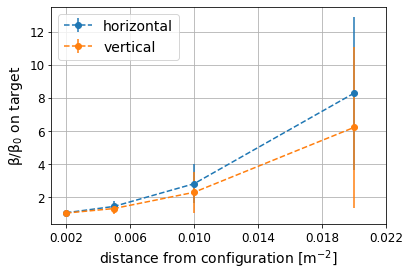} %
   \caption{Left plot: Average and maximum change of the $\beta$ function on target for configurations in a distance of $\rm 0.002~m^{-2}$ from the configurations found by the minimization procedure. Right plot: mean value of $\beta/\beta_{0}$ for various $\rm r_{21}$. }
   \label{fig:r21_investigation}
\end{figure}

The left plot of Figure \ref{fig:r21_investigation} shows the distribution of mean and maximum values 
of $\beta/\beta_0$ ratio on target for $r_{21} = \SI{0.002}{\per\square\meter}$. 
The right plot shows the evolution of the $\beta/\beta_0$ as a function of 
investigated distance $r_{21}$ from the original configuration.
From this analysis we can conclude that the typical size of a configuration space occupied by a configuration is
about \SIrange{0.002}{0.005}{\per\square\meter}, however closer analysis reveals much larger structures.

The various characteristic lengths are summarized Table \ref{tab:lengths}.

\begin{table}[!hbt]
   \centering
   \caption{Overview of various characteristic lengths (measured as Euclidean distances). "fn" stands for "farthest neighbor" and "nn" stands for "nearest neighbor" in configuration space. "initial-final" denotes the distance of corresponding initial and final configuration pairs. Distances are given in units of \si{\per\square\meter}.}
   \begin{tabular}{lrrrrrrr}
    \toprule
    {} &  mean &   std &   min &   \SI{25}{\percent} &   \SI{50}{\percent} &   \SI{75}{\percent} &   max \\
    \midrule
    fn-initial & 1.275 & 0.078 & 1.023 & 1.222 & 1.274 & 1.328 & 1.571 \\
    nn-initial & 0.410 & 0.044 & 0.243 & 0.381 & 0.410 & 0.439 & 0.634 \\
    fn-final   & 1.163 & 0.090 & 0.899 & 1.100 & 1.160 & 1.225 & 1.507 \\
    nn-final   & 0.296 & 0.042 & 0.136 & 0.268 & 0.295 & 0.323 & 0.479 \\
    initial-final & 0.456 & 0.124 & 0.038 & 0.368 & 0.446 & 0.535 & 0.992 \\
    \bottomrule
    \end{tabular}
   \label{tab:lengths}
\end{table}

%%%%%%%%%%%%%%%%%%%%%%%%%%%%%%%%%%%%%%%%%%%%%%%%%%%%%%%%%%%%%%%%%%%%%%%%%%%%%%%%
\section{Configuration space dimensionality reduction}
\label{sec:DimensionalityReduction}

The application of various constraints confines the valid configurations to a region of the configuration space that potentially needs fewer dimensions to be described (rather than the full number of 21 dimensions of the original configuration space). Since the structure of this region is neither identified nor apparent, a dedicated method for identification of the corresponding dimensionality is needed. We are using the method presented in \cite{number-of-intrinsic-dimensions} which relies solely on the distances of the two nearest neighbors for each data point. This has the advantage that the bias due to curvature or density variations is reduced in the estimate.

Table \ref{tab:intrinsic-dimensions} shows the resulting estimates for the three data sets. These indicate that further constraining the beta function along the beamline or at the target location, beyond the values of \SI{500}{\meter} and \SI{1}{\meter}, does not decrease the number of intrinsic dimensions. Hence we use the \dataA data set as a representative for the further analysis.

\begin{table}[!hbt]
   \centering
   \caption{Estimation of number of intrinsic dimensions for the different data sets.}
   \begin{tabular}{l|cc}
       \toprule
       \textbf{data set} & \multicolumn{2}{c}{\textbf{intrinsic dimension}}  \\
        & \textbf{initial} & \textbf{final} \\
       \midrule
           \dataA & 16.23 & 12.93 \\
           \dataB & 16.27 & 12.07 \\
           \dataC & 16.05 & 12.62 \\
       \bottomrule
   \end{tabular}
   \label{tab:intrinsic-dimensions}
\end{table}

%%%%%%%%%%%%%%%%%%%%%%%%%%%%%%%%%%%%%%%%%%%%%%%%%%%%%%%%%%%%%%%%%%%%%%%%%%%%%%%%
\section{Configuration space microstructure}
\label{sec:Microstructure}

The accuracy of the power supplies used in HEST is about 100 ppm and the precision is about 200 ppm. 
%(although the calibration could be performed to improve precision). 
Therefore the total relative uncertainty of magnet current setting is about $\delta=\Delta I/I = \num{3d-4}$ \cite{AStaf} which also applies to the relative uncertainty of each \kL-value. The potential difference between theoretical and actual setting in terms of distance in \kL-space depends on the \kL-values and has the following upper bound: 

%smallest distance between two configurations which can be distinguishable set is\footnote{Control system may impose additional limits on the setting uncertainty, but in this case the effect from power converters is dominant.}:

\begin{equation}
    \Delta k_{1, tot} = \delta\cdot\sqrt{6\cdot (\max|k_{1, QPK}|)^2 + 15\cdot (\max|k_{1,QPL}|)^2} = \SI{7.5d-4}{\per\square\meter}
\end{equation}

In order to investigate the configuration space microstructure, for each configuration in data set \dataA, a set of \num{5000} additional configurations was generated inside the 21-dimensional ball with radius \SI{0.001}{\per\square\meter} around that original configuration. These additional configurations were filtered according to the matching constraints and those that fulfill the constraints, in the following called \textit{leaf configurations}, are used for further analysis. These leaf configurations are the result of pure Monte Carlo sampling without any matching procedure. Hence they do not reflect any properties of the previously used LMDIF matching algorithm.

\begin{figure}[htb]
   \centering
   \includegraphics[width=.45\textwidth]{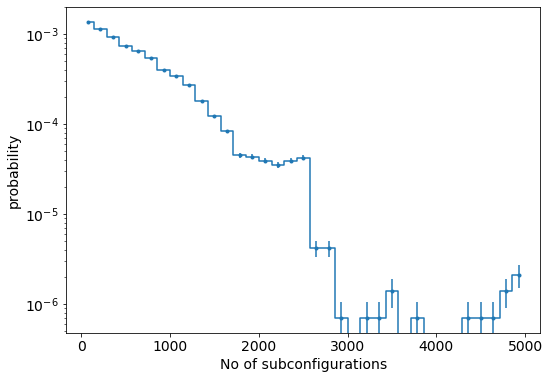}   
   \includegraphics[width=.45\textwidth]{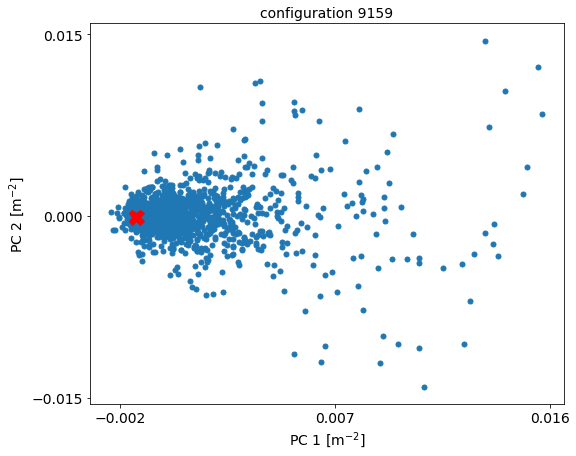}  
   \caption{Left: Number of leaf configurations found within the 21-ball of radius \SI{0.001}{\per\square\meter} around each configuration in the data set \dataA. Right: Example of a distribution of leaf configurations corresponding to a particularly large region of valid configurations. The two-dimensional distribution is obtained by projecting onto the plane of largest variance (principal components). The seed configuration is marked with a red "x".}
   \label{fig:disp_size}
\end{figure}

The distribution of the number of leaf configurations is shown on the left plot of Figure \ref{fig:disp_size}. The configuration space structure at this scale is very rich. In some areas the sampling method found no leaf configurations, what means that the original configuration is vulnerable to small errors in quadrupoles setting. On the other hand, there are a few areas which are filled with many leaf configurations and these areas are tolerant towards quadrupole errors.

An example of a region with a large number of good configurations is shown on right plot of Figure \ref{fig:disp_size}. The data for this plot was obtained using the above mentioned technique of probing the configuration space around each seed configuration in \dataA, with additional sampling of consecutive shells, each with the same thickness of \SI{0.001}{\per\square\meter} and containing \num{5000} samples, until no leaf configurations are found anymore. Since the volume of each shell increases with power 21, the intersection and thus the number of leaf configurations per shell, will decrease in case the probed region of potentially valid configurations locally spans less than 21 dimensions. This implies that even though the number of leaf configurations goes to zero, there is no evidence that the region of valid configurations is bounded at the same level in configuration space, just that the probability to sample a configuration in the intersection of that region with the hypershell decreases accordingly.
Nevertheless the thus obtained data gives an idea about the spatial distribution of leaf configurations around particularly good seed configurations.

% dispersion-size plot from many_shapes notebook

% dispersion-size plot from 5a notebook
%\begin{figure}[htb]
%   \centering
%   \includegraphics[width=.45\textwidth]{plots/Dx_cof_size_v0.png}   
%   %\includegraphics[width=.45\textwidth]{plots/pca_density_map_v2.png}   
%   \caption{Dispersion versus configuration size.}
%   \label{fig:disp_size}
%\end{figure}

%%%%%%%%%%%%%%%%%%%%%%%%%%%%%%%%%%%%%%%%%%%%%%%%%%%%%%%%%%%%%%%%%%%%%%%%%%%%%%%%
% jupyter 2
\section{Principal Component Analysis}
\label{sec:PCA}

Principal component analysis allows to find the main degrees of freedom of the studied system and thus to potentially reduce dimensionality of the configuration space.
The aspect of dimensionality reduction is useful for visualization and processing of high-dimensional data sets.

\begin{figure}[htb]
   \centering
   \includegraphics[width=.45\textwidth]{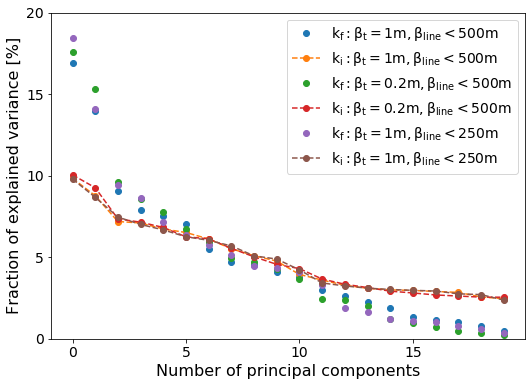} %jupyter 02
   \includegraphics[width=.45\textwidth]{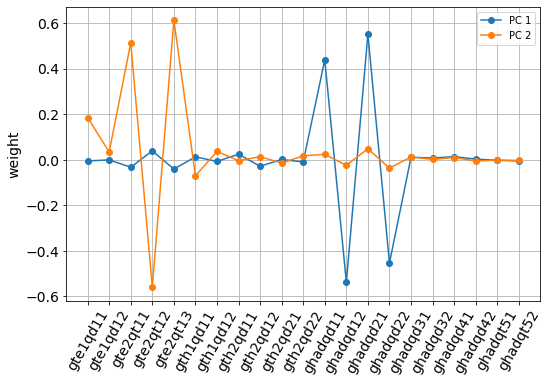}   
   \caption{Left: Fraction of the variance in the data explained by the given principal components - the first two principal components explain about \SI{30}{\percent} of the variance. Right: Composition of the first two principal components for the \dataA data set (for other data sets similar results are obtained). The horizontal axis shows the name of the quadrupole magnet along the beamline and the vertical axis shows the weight associated with corresponding \kL-value. }
   \label{fig:pca_components}
\end{figure}

The optimal number of principal components can be found using method from \cite{Minka}. The method, applied to the data set \dataA, shows that the first two components are responsible for about \SI{30}{\percent} of the total variance in the data and the remaining components vary significantly less, as illustrated in the left plot of Figure \ref{fig:pca_components}. 
%In this plot, the results for three different sets of constraints are shown. 
%The do not differ significantly, but it is worth noting that 
Imposing a stronger constraint on target focusing makes the first two principal 
components more pronounced, i.e. responsible for a larger fraction of the total variance.

The weights associated to the first and second principal components, here called PC1 and PC2, are shown on the right plot of Figure \ref{fig:pca_components}.
The first principal component mainly consists of the contribution of the two quadrupole doublets $\textrm{GHADQD}(11|12)$ and $\textrm{GHADQD}(21|22)$ while the second component corresponds to the quadrupole triplet 
$\textrm{GTE2QT}(11|12|13)$ at the beginning of the beamline. 
%It should be noted that the degree of freedom to modify the triplet strenght is overestimated here as in the 
%reality, in the lack of the appropriate steerers, it is used to steer the extracted beam

\begin{figure}[htb]
   \centering
   \includegraphics[width=.45\textwidth]{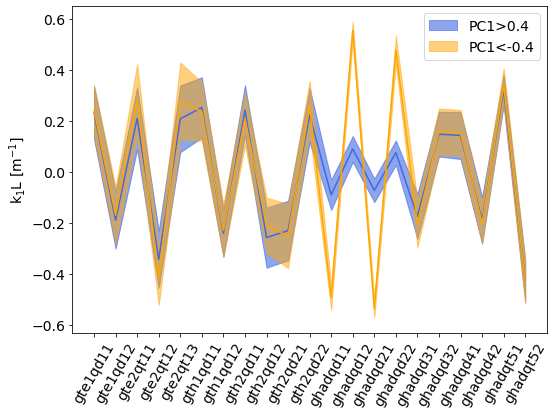}
   \includegraphics[width=.45\textwidth]{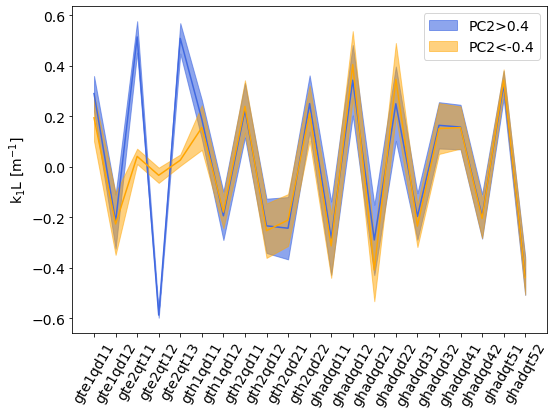}
   \caption{Visualization of configurations corresponding to large magnitudes of the first (left plot) and second (right plot) principal components.}
   \label{fig:pca12_composition}
\end{figure}

In order to better understand the meaning of the principal components, the plots of Figure \ref{fig:pca12_composition} shows the \kL - values
of all quadrupole magnets for those configurations that expose extreme values of PC1 and PC2. Negative values of PC1 correspond to strong focusing in the GHADQD-zone and positive values of PC2 correspond to strong focusing in GTE2QT-zone.
%This behaviour seems at first counter-intuitive, but the principal component method used in this study, applies centering for each feature before running SVD algorithm.

\begin{figure}[htb]
   \centering
   \includegraphics[width=.45\textwidth]{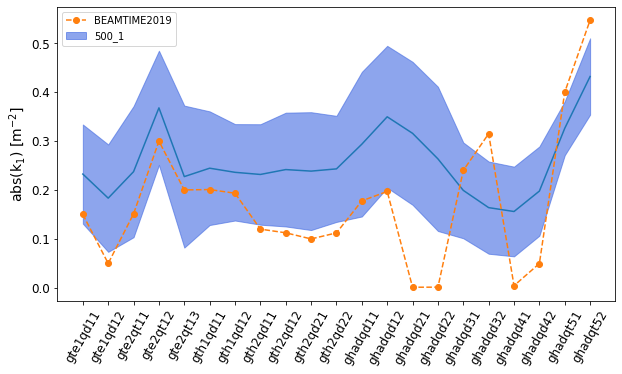} 
   \caption{ Mean and standard deviation of \kL-values along the beamline. The BEAMTIME2019 configuration, used in operation, is shown as well.}
   \label{fig:k1meansigma}
\end{figure}

The Figure \ref{fig:k1meansigma} shows the mean value as well as standard deviation of the absolute \kL-values along the beamline. The final focusing magnets have large \kL-values but rather small spread of the values. 
%The area which dominates first principal component is marked with a red box. 
The plot also shows the BEAMTIME2019 optics configuration which is commonly used in daily operation.

%tend to set the second {\it HAD} quadrupole doublet to very small values. 

%For comparison, in the Principal Components transformation used for BEAMTIME2019 setting, the
%very small values of second {\it HAD} quadrupole dublet \kL~ values are boosted by the fact that the mean values,
%used for centering, calculated over the whole sample of settings, are high.

\begin{figure}[!htb]
   \centering
   \includegraphics[width=.45\textwidth]{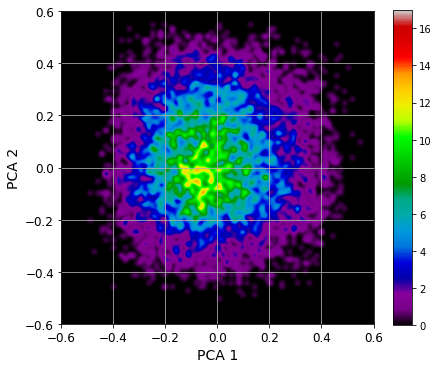} % 02
   \includegraphics[width=.45\textwidth]{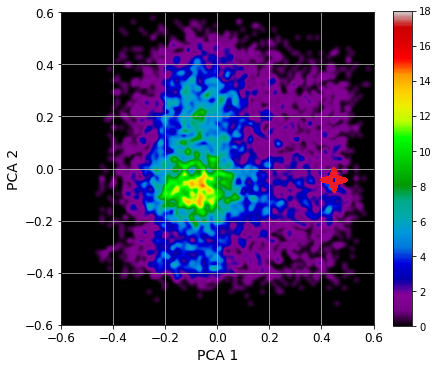}   %jupyter 02 -> move
   \caption{Density of models in \textit{PC-space} for the initial (left plot) \kL-value settings and the ones after the matching procedure converged (right plot).
   The star on the right plot shows the position of ion optics settings currently used in operation (BEAMTIME2019, see Figure \ref{fig:HADES_optics2019}).}
   %. Right plot: relative difference, blue areas show attractors in PCA space while red ones are where the models tend to avoid during matching process.}
   \label{fig:pca_mountains}
\end{figure}

Figure \ref{fig:pca_mountains} shows a density map of all configurations projected onto the plane corresponding to the first two principal components. This is referred to as \textit{PC-space} in the following.
The left plot shows a density map corresponding to initial \kL-values which were used as starting points for the matching procedure; only those values that could be successfully optimized are included. The right plot shows the corresponding \kL-values after the matching procedure converged.
The models are spread, but clearly the preferred values are for small negative first and second principal components. 
More than \SI{50}{\percent} of the configurations lie in the area defined by \hbox{$-0.2 < \textrm{PC1, PC2} < 0.0$}.

%which has actually a complex structure as shown on a zoomed plot in Figure \ref{fig:pca_zoom}.

%One must remember that this is an approximate configuration space, because the first two components explain only about 30\% of variance between models.

%The left plot of Figure \ref{fig:pca_zoom} contains zoom to the peak present in the PCA configuration space. This zoom shows additional small-scale structures. 

%%%%%%%%%%%%%%%%%%%%%%%%%%%%%%%%%%%%%%%%%%%%%%%%%%%%%%%%%%%%%%%%%%%%%%%%%%%%%%%%
% jupyter 12
%\section{configuration space volume - remove} - calculation was wrong

%%%%%%%%%%%%%%%%%%%%%%%%%%%%%%%%%%%%%%%%%%%%%%%%%%%%%%%%%%%%%%%%%%%%%%%%%%%%%%%%
% jupyter 3,4
\section{Clustering}
\label{sec:Clustering}

The goal of clustering is to investigate if the optics configurations can be divided into groups of common features.
Several algorithms were tested, but the only interesting results were obtained using the k-means algorithm.
%in 21-dimensional \kL - parameter space.
This algorithm performs a partitioning of the data, that is it does not find an optimal number of clusters but it divides the data into a predefined number of partitions.

\begin{figure}[htb]
   \centering
   \includegraphics[width=.45\textwidth]{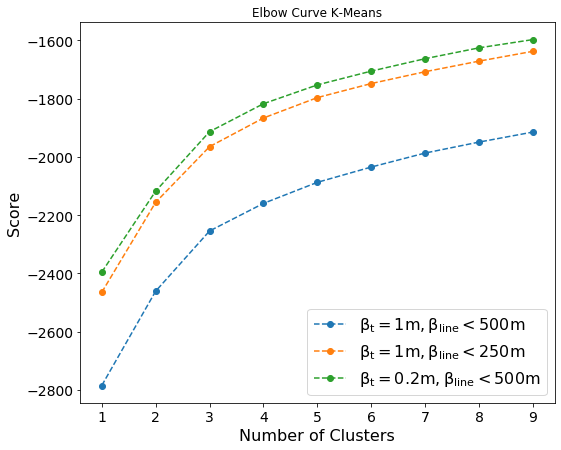}
   \includegraphics[width=.45\textwidth]{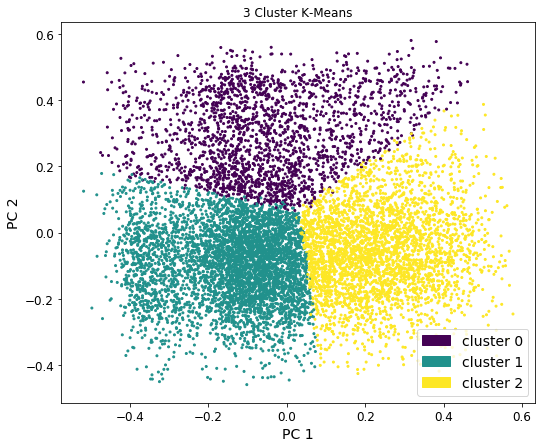}
   \caption{Left: elbow curve for k-means algorithm. Right: cluster coverage in principal component space.}
   \label{fig:k-means-elbow}
\end{figure}

The elbow curve, shown on the left plot of Figure \ref{fig:k-means-elbow}, 
can be used to estimate the optimal number of clusters. 
The choice of number of clusters is less straightforward than the choice of minimum number of Principal Components. 
Visually the number of three clusters
seems to be a better choice than two or four.  
The subsequent results were obtained requesting three clusters.

The right plot of Figure \ref{fig:k-means-elbow} shows division of the PC component configuration space into 3 clusters. 
Those clusters are clearly distinguishable and contain respectively: \SI{21}{\percent}, \SI{47}{\percent} and \SI{32}{\percent} of configurations.
If two clusters are requested, the partitioning reveals a clear distinction between negative and positive values of the second principal component (PC1), ie. configurations are split between strong and weak focus in GTE2QT segment.
If more than three clusters are chosen, the resulting partitions overlap in the PC1-PC2 projection of the configuration space.

The clustering algorithm is applied in \kL-space, but the same results are obtained when it is applied in PC-space. It means that the features detected by the algorithm are present in the first two Principal Components and not in other components.

%\begin{figure}[htb]
%   \centering
%   \includegraphics[width=.45\textwidth]{plots/2_Cluster_K-Means.png}
%   \caption{Effect of finding 2 or 3 clusters in 21-D parameter space of k-values.}
%   \label{fig:k-means-cluster}
%\end{figure}
 
%It shows that the beamline settings can be seen as a linear mixture (angle) of the two Principal Component factors.
%In case of two clusters, the K-Means procedure finds that having net focusing or net defocusing
%in ghadqd11-22 zone. 

%Clusters are not very "accented" (no big distances in PCA space).

%16-clustering.ipynb
\begin{figure}[htb]
   \centering
   \includegraphics[width=.45\textwidth]{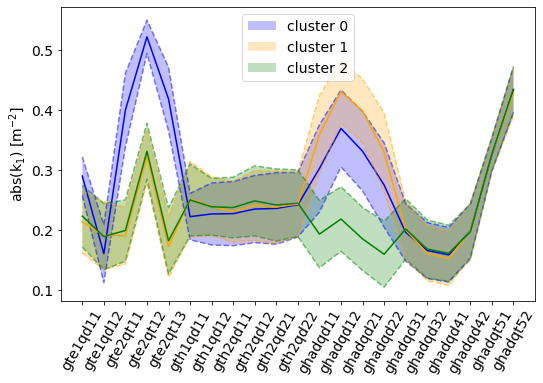}
   \includegraphics[width=.45\textwidth, height=.312\textwidth]{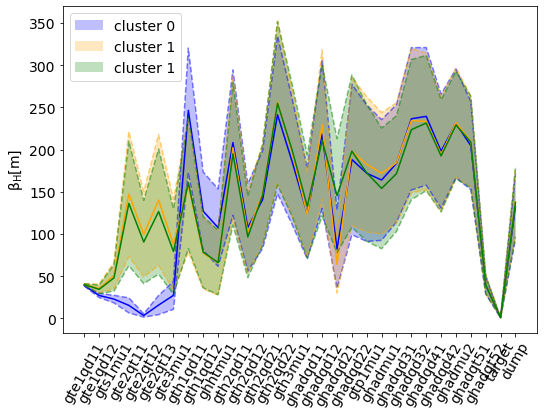} % jupyter 03
   \caption{Left: the distribution of quadrupoles k-values along the beamline for 3-cluster analysis. For better visibility the absolute value of k-values is displayed. Right: horizontal beta function along the line for the three clusters. The line represents mean value and the band represent half of standard deviation.}
   \label{fig:k-val-cluster}
\end{figure}

The left plot of Figure \ref{fig:k-val-cluster} presents the distribution of \kL-values and the horizontal optics functions for the three clusters.
It is interesting to note that the unsupervised algorithm, which is k-means clustering, finds three main strategies
for ion optics on the HADES beamline:
\begin{itemize}
\item cluster 0, with strong focus in GTE2QT segment, %and medium focus in HAD zone,
\item cluster 1, with weak focus in GTE2QT segment and strong focus in GHADQD segment,
\item cluster 2, with weak focus in GTE2QT segment and weak focus in GHADQD segment.
\end{itemize}

%A closer investigation of the Figure \ref{fig:k-val-cluster} leads to a conclusion that clusters 2 and 3 are not as well separated as cluster 1. So the main two strategies resulting from the minimisation procedure is strong and weak focus in TE2 region,
%which is defined by the second principal component (PCA2).

%\vspace{10mm}
%%%%%%%%%%%%%%%%%%%%%%%%%%%%%%%%%%%%% MAYBE
%\include{StatOptics_PhaseAdvance}

%%%%%%%%%%%%%%%%%%%%%%%%%%%%%%%%%%%%%%%%%%%%%%%%%%%%%%%%%%%%%%%%%%%%%%%%
% jupyter 06 - not anymore
\section{Stability of optics configurations}
\label{sec:Robustness}

The stability of an optics configuration can refer to two different aspects. 
One aspect is the change of the beta functions along the beamline and at the target location as a function of change of the Twiss parameters at the entrance of the beamline. Such a shifting of lattice parameters can lead to an increase of beam spot size at the target and hence it is desirable that a configuration is \textit{robust} against such shifting.

The other aspect is concerned with quadrupole gradient errors. Small changes in the \kL-values might lead to an increase of the beam spot size at the target location and hence it is desirable that a configurations is \textit{tolerant} towards such gradient errors.

Here we define the \textit{robustness} and \textit{tolerance} of configurations as follows. The robustness score is given by the formula:

\begin{equation}
    \textrm{Robustness} = \sqrt{\max\left(\frac{\Delta\beta_{h, \textrm{target}}}{\beta_{h, \textrm{target}}}, 0\right)^2 + \max\left(\frac{\Delta\beta_{v, \textrm{target}}}{\beta_{v, \textrm{target}}}, 0\right)^2}
\end{equation}

The $\Delta\beta_{(h,v), \textrm{target}}$ is the result of a shifting of the Twiss parameters at the entrance of the beamline. The $\max$ part ensures that only an increase in beta functions is taken into the account and hence a robustness score of zero indicates that the configuration is robust against the shifting, i.e. it will not increase the beta function at the target location. A robustness score greater than zero indicates an increase in beta function by a corresponding magnitude.

The tolerance of a configuration towards quadrupole gradient errors can be assessed via the number of valid configurations that are found in a ball with corresponding radius around that configuration. This is similar to the sampled leaf configurations from Section \ref{sec:Microstructure}, where the distribution of tolerance is presented in Figure \ref{fig:disp_size}. Here we define the tolerance of a configuration as the fraction of leaf configurations:

\begin{equation}
    \textrm{Tolerance} = \frac{N_{leaves}}{N_{samples}} \hspace{1cm} \textrm{within } R = \SI{0.001}{\per\square\meter}
\end{equation}

\begin{figure}[htb]
   \centering
   \includegraphics[width=.45\textwidth]{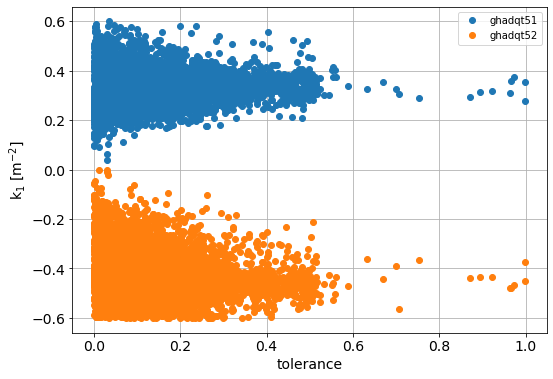}
   \caption{Strength of final focusing quadrupoles in function of tolerance.}
   \label{fig:k1ff_tolerance}
\end{figure}

Figure \ref{fig:k1ff_tolerance} presents a particular and non-trivial property of optics configurations: the most tolerant configurations have the \kL-values of the two last focusing quadrupoles close to their average value. Therefore, in order to speed-up the search for high-tolerance configurations, one could restrict the \kL-values of these magnets.

%16-clustering.ipynb

%refers to the configuration space volume of a configuration - a bigger volume corresponds to configuration more immune to gradient errors of quadrupole magnets.

% In this section beta function changes due to k-values errors of the quadrupole magnets or to  magnets misalignment. Here we chose a simpler approach, motivated by the characteristics of the slow extraction process.

The HADES experiment requires slowly extracted beams and the optics of the synchrotron change during the quadrupole-driven slow extraction process which causes the lattice parameters at the entrance of the transfer line to change too.
According to dedicated optics calculations \cite{SSorge} the values of $\beta_{h}$, $\beta_{v}$ 
and horizontal dispersion $D_{h}$ at the entrance of the beamline vary during the spill by \SI{3}{\percent}, \SI{1}{\percent} and \SI{3.5}{\percent} respectively. 
For the following analysis we recompute the beamline lattice functions with modified values at the entrance of the beamline
and observe the change of beta function at the experimental target location for each of the configurations in the data set \dataA. The change of beta functions at the target location is denoted with $\Delta\beta_{h}$ and $\Delta\beta_{v}$.

%We use these values to investigate the change of the beta functions on the target ($\beta_{t,H}$, $\beta_{t,V}$).
%Basically we investigate $\rm \delta\beta_t / \delta\beta_b $. 
%The range of the changes was divided into 10 sub-ranges to investigate the linearity of the derivative. 
%It was found that, in this range of optics changes at the beginning of the beamline, the change at the end of the beamline is linear.

%The stability of the obtained settings were tested with respect to change of the beam parameters
%at the beginning of the beamline. The paramteres:  The range of parameters change were indicated by the study
%of the optics change during the slow extraction process \cite{SSorge}. The range was divided into 10 steps and, for each step, the optics functions along the beamline were %calculated.

\begin{figure}[htb]
   \centering
   \includegraphics[width=.45\textwidth]{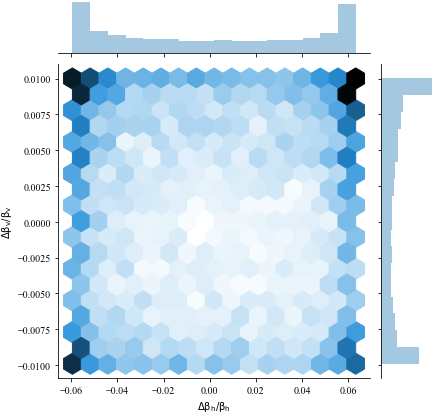}
   \includegraphics[width=.45\textwidth]{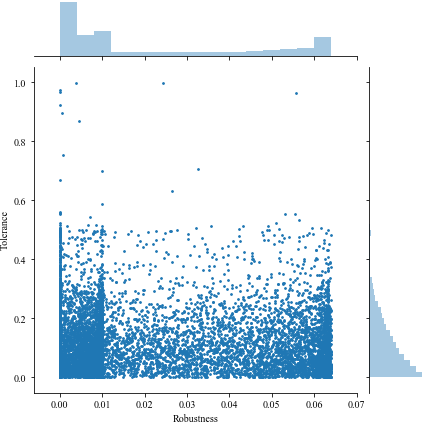}  %    
   \caption{Relative change of the beta functions on experimental target due to variation of the beam parameters at the beginning of the beamline.}
   \label{fig:stability}
\end{figure}

As the beamline lattice is mostly linear, the expected change of lattice parameters at the end of the line is of the same order of magnitude as the variation at the entrance of the beamline.
The relative changes of beta functions at the target location are shown on the left plot of Figure \ref{fig:stability}.

The vertical change is about \SI{30}{\percent} of the horizontal one, as expected from the changes imposed at the entrance of the beamline and the fact that
part of the beamline is tilted, therefore leading to coupling of the two planes. 

%This is illustrated in Figure \ref{fig:stability}, where distribution of the relative change of beta functions on the experimental target is shown. 
The distribution has clear maxima - deviation of the optics function on the target is non-zero, however a small population of configurations
lie in a wide minimum of the configuration space, where they seem quite independent on the variation of beam parameters at the beginning of the beamline.
%It is interesting to note that this distribution is centered around zero, what means that for many of the investigated models a better focusing on the target is possible.

% investigate also beam position?
% which clusters are most robust?

The results show that \SI{22}{\percent} of the models are robust against a shifting of the lattice functions with a few models leading even to a significant decrease of the beta function at the target location in both planes.
% For the moment we have not found correlations between robustness and position in the PC configuration space or clusterization.

%About how much a change of betas and dispersion affect the optics (sum or maximum of delta betas, delta beta on target)

%\vspace{10mm}

%%%%%%%%%%%%%%%%%%%%%%%%%%%%%%%%%%%%%%%%%%%%%%%%%%%%%%%%%%%%%%%%%%%%%%%%%%%%%%%%
%\include{sub/matching_convergence.tex}

%\vspace{10mm}

%\include{StatOptics_NeuralNetworks}

%%%%%%%%%%%%%%%%%%%%%%%%%%%%%%%%%%%%%%%%%%%%%%%%%%%%%%%%%%%%%%%%%%%%%%%%%%
% jupyter 15
\section{Ion optics choice}
\label{sec:OpticsChoice}

\begin{figure}[htb]
   \centering
   \includegraphics[width=.45\textwidth]{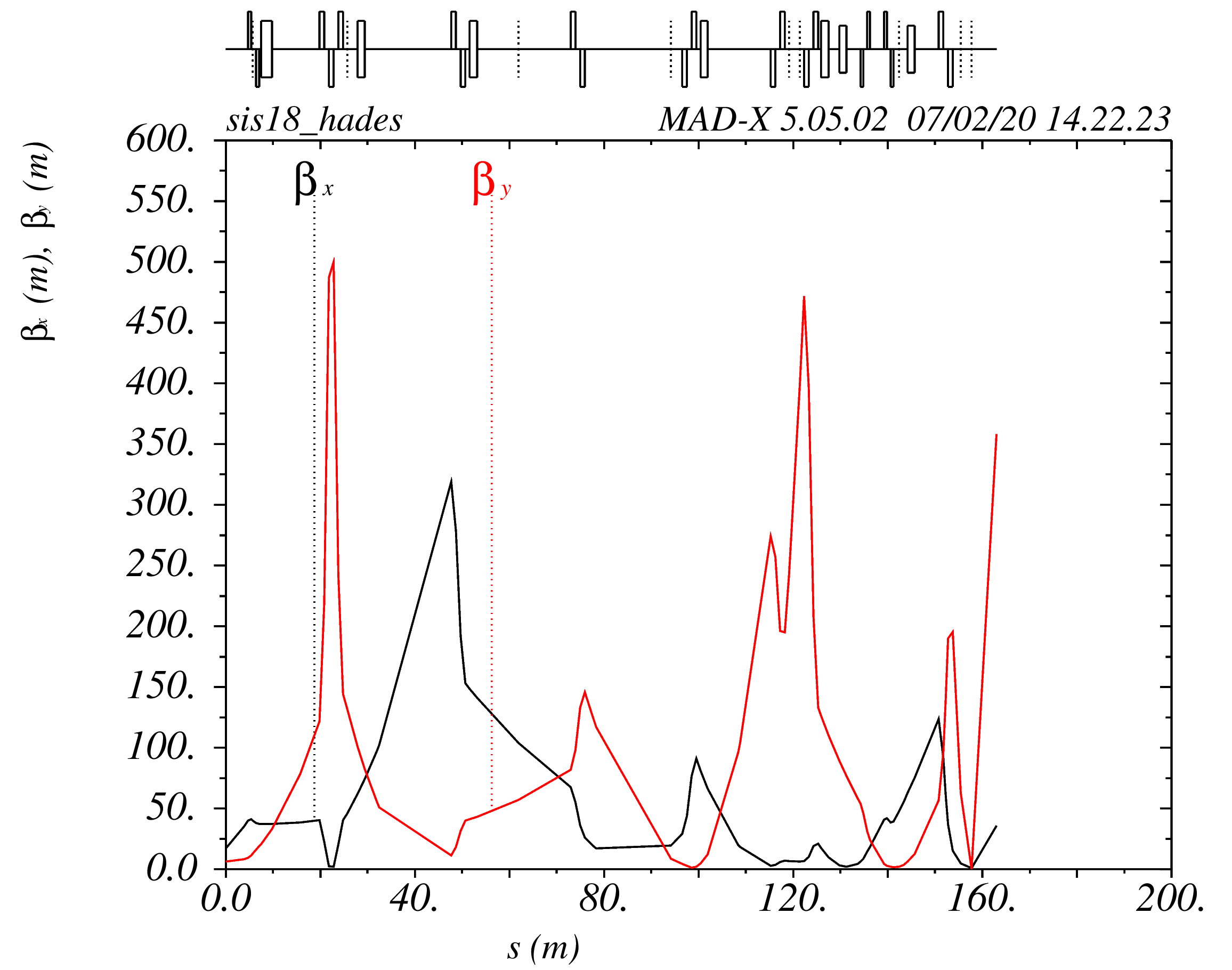}    %    
   \includegraphics[width=.45\textwidth]{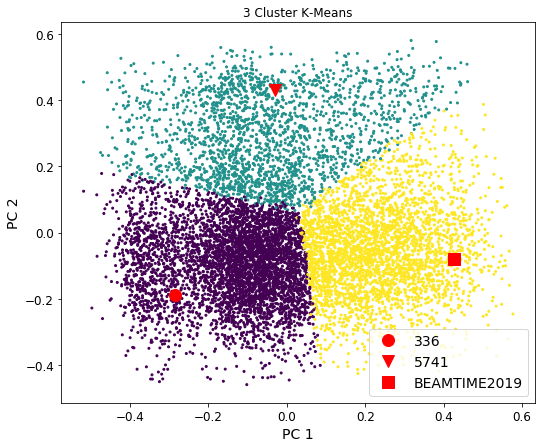}    %    
   \caption{Left: optics functions for configuration \#5741. Right: location of the discussed configurations in principal component space.}
   \label{fig:ionoptics2}
\end{figure}

An interesting observation is that most of the historically used ion optics settings of the
beamline are located in same region of PC-space, at about $\rm PC1 = (0.42,0.44)$ and $\rm PC2=(-0.05,0.05)$ (see Figures \ref{fig:pca_mountains}, \ref{fig:ionoptics2}). In this region the focus of GHADQD magnets is weak and for GTE2QT magnets it is moderate (belonging to cluster 2).
%This area, as seen in right plot of Figure \ref{fig:euclidean_distance} is an attractor, so one can expect this setting is in a kind of local minimum (or optimum), however it is quite peripheral.

As mentioned before, there is no single best solution for ion optics of a multi-purpose beamline. However, the configurations found in the exhaustive scan of the available configuration space reveal various levels of tolerance towards quadrupole errors, robustness with respect to shifting of lattice functions and are characterized by various levels of dispersion at the target location.

\begin{figure}[htb]
   \centering
    \includegraphics[width=.45\textwidth]{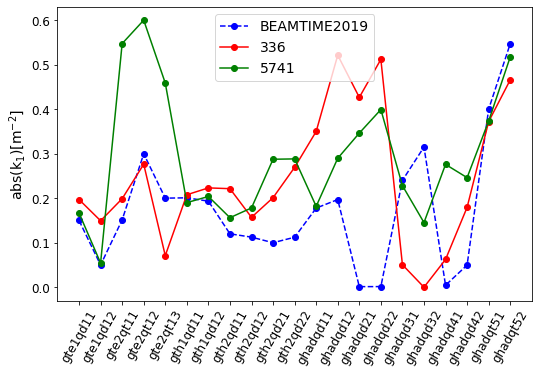}   
   \includegraphics[width=.45\textwidth]{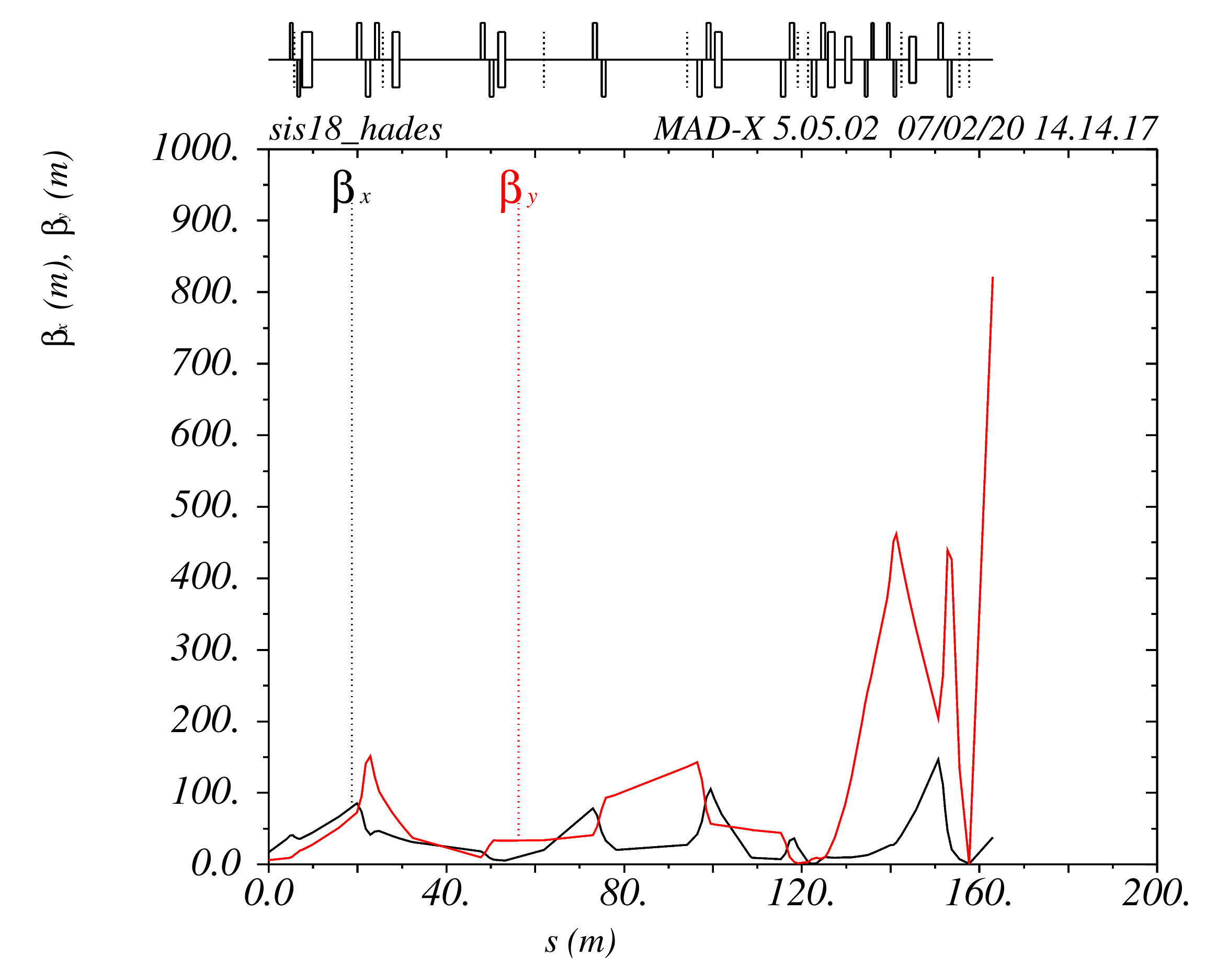}    %    
   \caption{Left: comparison of \kL-values for operational ion optics settings with two new settings proposed as an outcome of the study. Right: optics functions for configuration \#336.}
   \label{fig:ionoptics1}
\end{figure}

Two particular configurations have been investigated as a potential substitute for historically used settings. The first, no. 336, has a robustness score of zero (full robustness) and maximal tolerance. The second one, no. 5741, is chosen with similar criteria but selecting only from configurations which have very small horizontal dispersion on the target: $D_x < \SI{0.1}{\meter}$.
 These configurations are shown in Figures \ref{fig:ionoptics1} and \ref{fig:ionoptics2}.

%15-new-optics.ipynb

Right plot of Figure \ref{fig:ionoptics2} suggests that good configurations are spread across the {\it PC-space}, without any particular regularity or preferences.

%%%%%%%%%%%%%%%%%%%%%%%%%%%%%%%%%%%%%%%%%%%%%%%%%%%%%%%%%%%%%%%%%%%%%%%%%%%%%%%%
\section{Conclusions}

%A statistical analysis of a beamline properties is presented here.
The analyzed beamline is designed to be flexible as other beamlines bifurcate from it and the main experiment HADES, which is situated at the end of this line, operates in various modes.
The methodology of the study relies on generation of large amounts of configurations spread in the available configuration space and by executing the matching procedure with constraints on focusing beam on the target and keeping beam envelope within the beamline acceptance.

\newpage

The following conclusions are the highlights of the study:

\begin{itemize}
    \item When stronger constraints are imposed on the target focusing, more quadrupoles along the beamline are employed to meet the constraint, what is expressed by selection of particular values of a total phase advance.
    \item The configuration space of matched optics fills a small, but continuous region of the total possible configuration space.
    \item However, this region is not uniform: there are configurations which are tolerant to change of \kL-values and others which are not; This is expressed as a variation of  density of matched configurations.
    \item Principal Component Analysis reveals that two beamline sections, one at the beginning and one in the middle have most variance while setting of other magnets is more constrained.
    \item Partitioning the available configurations shows that three approaches to construction of beamline optics can be distinguished, based on values of the first two Principal Components.
    \item Configurations have also varying tolerance to the change of the initial twiss parameters.
    \item Neither tolerance to varying initial twiss parameters nor robustness of the configurations were found to favour some particular region of configuration space.
    \item A choice of new optics configurations can be done based on selection of parti\-cu\-lary robust and tolerant configurations.
\end{itemize}

The presented analysis allows for investigation of the possible ion optics settings of a beamline. It reveals what types of optics are possible and gives indications about
sensitivity of the optics to various errors, like uncertainty of beam parameters at the entrance 
to the beamline or quadrupole errors. 

In the future developments, it is planned to check other matching procedures, especially gradient-free ones and to perform more precise studies of configuration space microstructure.

%We plan to continue this study using other minimization methods, for instance gradient-free optimizers.
%We plan as well to investigate other estimations of robustness of the settings as well as size of the 
%'attractors' (ie. stability areas).

%%%%%%%%%%%%%%%%%%%%%%%%%%%%%%%%%%%%%%%%%%%%%%%%%%%%%%%%%%%%%%%%%%%%%%%%%%%%%%%%

\end{document}